# The Hydrogen Atom of Bohr II: Mean Life Time of an Ensemble of Excited Hydrogen Atoms

P. S. Kamenov

Similar to the results in [1, and paper 9902012 /xxx.lanl.gov], we established in [2] that the "own life time" (t) of one single excited atom (in state (n)) depends exactly on the energy difference ($\Delta E$) [2]:

(1) $$t = \frac{3\hbar R}{2(\Delta E)^2 n^2}$$

The own life time (t [s]) is determined by the exact energy of excitation ($\Delta E = |E_n - E|$), the constants of Planck ($\hbar$) and Rydberg (R), and the principle quantum number (n) of the excited state. This time cannot be measured experimentally (except in the case shown in [1] for resonant Mossbauer transitions in nuclei). Experiments with hydrogen measure only the mean life time of an ensemble of excited atoms.

Further we attempt to find the mean life times ($\tau_n$) (for different excited states) of an ensemble of hydrogen atoms and compare these life times with reference data. Let us assume that $N_0$ [cm$^{-3}$] atoms are irradiated by a flux of photons with uniform energy distribution $\Phi(E) = \Phi_0$[cm$^{-2}$s$^{-1}$] = const. The effective cross-section of excitation is $\sigma_E$. Then, the activity which can be obtained is:

(2) $$\frac{dN}{dt}(t) = \Phi_0 \sigma_E N_0 \left(1 - \exp(-t/\tau_n)\right)$$

As it is well known, after irradiation is terminated, activity changes with time in the following way:

(3) $$\frac{dN}{dt}(t) = \Phi_0 \sigma_E N_0 \left(\exp(-t/\tau_n)\right)$$

On the other hand, the differential cross-section ($d\sigma_E$) is:

(4) $$d\sigma_E = \frac{\sigma_0 \Gamma_n dE}{4(\Delta E)^2 + \Gamma_n^2}$$

And then effective cross-section ($\sigma_E$) will be:

(5) $$\sigma_E = \frac{\pi \sigma_0}{2}$$

Substituting (5) in (3) we obtain the variation of activity with time after excitation:

(6) $$\frac{dN}{dt}(t) = \Phi_0 \frac{\pi \sigma_0}{2} N_0 \left(\exp(-t/\tau_n)\right)$$

Under the same conditions, but using the differential cross-section (4), we can find how activity $\frac{dN}{dt}(E)$ increases with irradiation time:

(7) $$\frac{dN}{dt}(E) = \frac{\Phi_0 N_0 \sigma_0 \Gamma_n dE}{4(\Delta E)^2 + \Gamma_n^2}\left(1 - \exp(-t/\tau_n)\right)$$

In order to derive an expression for this activity after the end of irradiation, from (1) we obtain the variation of the own life time (t) with energy:

(8)

$$dt = \frac{3\hbar R dE}{(\Delta E)^3 n^2}$$

Because of the symmetry of (1) with respect of energy, in the time interval (dt) decay the atoms in the two intervals $\Delta E$ on both sides of $E_n$:

(9) $$dt = \frac{3\hbar R dE}{(\Delta E)^3 n^2} + \frac{3\hbar R dE}{(\Delta E)^3 n^2} = \frac{6\hbar R dE}{(\Delta E)^3 n^2}$$

or

(10) $$dE = \frac{(\Delta E)^3 n^2 dt}{6\hbar R}$$

Substituting (dE) in (7) one can find the activity of hydrogen atoms (after irradiation):

(11) $$\frac{dN}{dt}(E) = \frac{\Phi_0 N_0 \sigma_0 \Gamma_n (\Delta E)^3 n^2 dt}{(4(\Delta E)^2 + \Gamma_n^2) 6\hbar R}$$

Thus we obtain two expressions for the activities: (11), depending on the energy of excitation (ΔE), and (6), depending on time (t). Using the previous results from [1], we require that the two activities (6) and (11) be equal:

(12) $$\frac{\Phi_0 N_0 \sigma_0 \Gamma_n (\Delta E)^3 n^2 dt}{(4(\Delta E)^2 + \Gamma_n^2) 6\hbar R} = \Phi_0 \frac{\pi \sigma_0}{2} N_0 (\exp(-t/\tau_n))$$

In the specific case when $\exp(-t/\tau_n) = 1/2$, then $\Delta E = \Gamma_n/2$, and the expression (12) becomes [1]:

(13) $$\frac{\Gamma_n^2 n^2 dt}{24\hbar R} = \pi$$

Hence, the natural width ($\Gamma_n$) for unit time interval (dt=1) can be calculated as:

(14) $$\Gamma_n = \frac{1}{n}\sqrt{24\pi\hbar R}$$

From the natural width ($\Gamma_n$) of level (n) it is easy to derive the mean life time of all excited atoms (at level n):

(15) $$\tau_n = \frac{\hbar}{\Gamma_n} = n\sqrt{\frac{\hbar}{24\pi R}}$$

Thus, for calculation of the full mean life time of an excited hydrogen level (n), only Rydberg's constant (R) and Planck's constant ($\hbar$) are needed. The corresponding decay constant (the spontaneous coefficient of Einstein) is $A_n = 1/\tau_n$.

***Comparison with reference data.*** In the numerous reference tables on hydrogen I found, to my great surprise, quite different values for $\tau_n$ (especially for high excited states). In the table below I quote the data from [3] (1966) and [4] (1986) and compare them with my calculations (formula 15, 1997). As it is seen, for the first excited state (n=2) the calculated $\tau_n$ is equal to 1.603x10⁻⁹ s, while in [3] this time is $\tau_n$=2.127x10⁻⁹ s and in [4] $\tau_n$=1.60x10⁻⁹ s. So, the result from the present calculations is in excellent agreement with reference data (for n=2). It is necessary to stress that my calculations fit better to the values in [4]. The differences between the values in [3] and [4] are greater than the differences between my calculations and the data in [4].

| | Data Sources | | | | | |
|---|---|---|---|---|---|---|
| | [3] (1966) | | [4] (1986) | | (1997) | |
| n | $\tau_n$, s | $A_n$, s$^{-1}$ | $\tau_n$, s | $A_n$, s$^{-1}$ | $\tau_n$, s | $A_n$, s$^{-1}$ |
| 2 | $2.12 \times 10^{-9}$ | $4.699 \times 10^{8}$ | $1.60 \times 10^{-9}$ | $6.25 \times 10^{8}$ | $1.603 \times 10^{-9}$ | $6.23 \times 10^{8}$ |
| 3 | $1.0 \times 10^{-8}$ | $1.0 \times 10^{8}$ | $3.94 \times 10^{-9}$ | $2.53 \times 10^{8}$ | $2.405 \times 10^{-9}$ | $4.15 \times 10^{8}$ |
| 4 | $3.3 \times 10^{-8}$ | $3.02 \times 10^{7}$ | $8.0 \times 10^{-9}$ | $1.24 \times 10^{8}$ | $3.2 \times 10^{-9}$ | $3.12 \times 10^{8}$ |

The values of $\tau_n$ from this work (1997) are closer to the values of data source [4] (1986). The difference between the data from [3] and [4] (for n>2) are impermissible.

**Differences between the data.** Here I will attempt to explain the great differences between reference data (for n>2). The experimental results are very good only for the firsts excited states... I think that the differences between reference data (for n>2) are due both to experimental difficulties and (mainly) to the wrong interpretation of the relation between Einstein's coefficients, which is explained in [2,5,6].

In [3] the transition probability for spontaneous emission from upper state k to lower state i, $A_{ki}$, is related to the total intensity $I_{ki}$ of a line of frequency $\nu_{ik}$ by

(16) $$I_{ki} = \frac{1}{4\pi} A_{ki} h \nu_{ik} N_k$$ (expression (1) on page ii of [3])

where h is Planck's constant and $N_k$ the population of state k. It was shown in [5,6] that this relation holds for transitions from any excited state k to the ground state i only. If (i) is also an excited state, then relation (16) must be:

(17) $$I_{ki} = \frac{1}{4\pi}(A_{ki} + \frac{g_i}{g_k}A_{ix})h\nu_{ik}N_k$$

where $A_{ix}$ is the full decay constant of level (i). Only when $A_{ix}=0$ (ground state) (17) coincides with (16). The same applies for the transition probability of absorption $B_{ik}$ and the transition probability of induced emission $B_{ki}$ in [3]:

(18) $$B_{ik} = 6.01\lambda^3 \frac{g_k}{g_i}A_{ki} \text{ (expr. (6), p. vi of [3])}$$

(19) $$B_{ki} = 6.01\lambda^3 A_{ki} \text{ (expr. (7), p. vi of [3])}$$

($\lambda$ is the wavelength in Angstrom units). When (i) is an excited state, these relations are also wrong. According to [5,6], these relations (in the same units as in [3]) will be:

(20) $$B_{ik} = 6.01\lambda^3 \left(\frac{g_k}{g_i}A_{ki} + A_{ix}\right)$$

(21) $$B_{ki} = 6.01\lambda^3 \left(A_{ki} + \frac{g_i}{g_k}A_{ix}\right)$$

It is also seen that if (i) is a ground state, $A_{ix} = 0$, these relations correspond to the relations in [3]. It is clear that even based on experimental results (when n>2), if processed using the inappropriate relations [3], $\tau_n$ can have wrong values.

Therefore, one can conclude that all reference data for transition probabilities in atoms (especially in hydrogen) must be critically examined and adjusted accurately.

### *Some inevitable conclusions.*

These simplest calculations show that Bohr's model of hydrogen is as adequate as real the field of L. de Broglie is. If the energy of excitation corresponds exactly to the conditions for a stationary state [2], the "own lifetime" of the excited level will be infinite (if there are no other external perturbations; such perturbations are very small in nuclear systems [1]). If the energy of excitation is different from that corresponding to the exact conditions for a stationary state, after some evolution of the excited state, the Coulomb field can change the state of the electron because the amplitude of de Broglie becomes zero and electron is no more in a potential well (the electron can emit a photon-soliton [7,8]). The own lifetime of the excited state depends on the energy difference between the actual excitation and exact condition for a "stationary" state (1). The difference between the exact energy (for stationary state) and the actual energy of excitation is this perturbation energy which is needed for spontaneous transition to a lower state (except the ground state, where de Broglie's "unitary field-particle" can not be destroyed). The main result of this work (and works [1,2]) is that excited states of the hydrogen atom [2] and the nucleus [1] decay after some exactly predictable time (t) according to (1). Decay is not an accidental event as it is believed by the majority of scientists (except Einstein who wrote that a weakness of the theory of radiation is that the time of occurrence of an elementary process is left to "chance" [9]).

This work was supported in part by the Bulgarian National Foundation for Scientific Research (No 534/1995).

Faculty of Physic
University of Sofia
1164 Sofia, Bulgaria